# Spin-orbit torque switching of perpendicular magnetization in ferromagnetic trilayers

Dong-Kyu Lee[1] and Kyung-Jin Lee[1,2,*]

[1] *Department of Materials Science and Engineering, Korea University, Seoul 02841, Korea*

[2] *KU-KIST Graduate School of Converging Science and Technology, Korea University, Seoul 02841, Korea*

* Correspondence and requests for materials should be addressed to K.-J. Lee (kj_lee@korea.ac.kr)

Abstract

In ferromagnetic trilayers, a spin-orbit-induced spin current can have a spin polarization of which direction is deviated from that for the spin Hall effect. Recently, magnetization switching in ferromagnetic trilayers has been proposed and confirmed by the experiments. In this work, we theoretically and numerically investigate the switching current required for perpendicular magnetization switching in ferromagnetic trilayers. We confirm that the tilted spin polarization enables field-free deterministic switching at a lower current than conventional spin-orbit torque or spin-transfer torque switching, offering a possibility for high-density and low-power spin-orbit torque devices. Moreover, we provide analytical expressions of the switching current for an arbitrary spin polarization direction, which will be useful to design spin-orbit torque devices and to interpret spin-orbit torque switching experiments.

**Introduction**

Current-induced magnetization switching is a basic working principle of magnetic random access memories (MRAMs). Perpendicular MRAMs[1,2], which store the magnetic information in a perpendicularly magnetized free layer, are of technological relevance because of better scalability than in-plane MRAMs. Current-induced magnetization switching schemes can be classified into two categories depending on the type of spin torque. One type is the spin-transfer torque (STT)[3,4], which utilizes a spin current polarized by the exchange splitting of the other ferromagnetic layer. In magnetic tunnel junctions consisting of two ferromagnets (FMs) separated by a thin insulator, one of the FMs supplies a spin-polarized current, which switches the other free FM layer[5,6]. For the STT switching scheme, the current-perpendicular-to-plane (CPP) geometry is inevitable because a charge current must pass through both FMs. The other type is the spin-orbit torque (SOT), which utilizes a spin current polarized by the spin-orbit coupling of a nearby normal metal (NM). The spin-orbit coupling enables the damping-like SOT through the bulk spin Hall effect[7-10] or the interfacial Rashba effect[11-13]. The SOT-induced perpendicular magnetization switching[14,15] occurs in the current-in-plane (CIP) geometry where a charge current flowing in the plane (i.e., $x$ direction) supplies a spin current flowing normal to the plane (i.e, $z$ direction), which in turn exerts a SOT on the free layer. In FM/NM bilayers, the spin polarization carried by a spin current is orthogonal to both directions of the charge-current flow ($x$) and the spin-current flow ($z$), and thus is aligned along the $y$ direction.

Compared to the STT switching, the SOT switching has important advantages due to the difference in the write-current path (i.e., CPP for STT switching vs CIP for SOT switching). The most important advantage of the SOT switching scheme is that the write-current path is separated from the read-current path, which naturally resolves the write-read interference[16].

Moreover, the device endurance is better for the SOT switching because large writing currents do not pass through an insulating layer. However, the SOT switching has two critical issues for device applications at the same time. One is that the switching current is too high. The other is that an additional symmetry-breaking field is required for the deterministic switching of perpendicular magnetization. As a result, much effort has been expended in realizing field-free SOT switching at a low current[17-23].

We note that both issues for the SOT-induced perpendicular magnetization switching originate from the fact that the spin polarization ($y$) of spin current is orthogonal to the equilibrium magnetization direction ($z$). Because of this orthogonal configuration, the SOT does not directly compete with the damping torque and, as a result, the switching current is independent of or less dependent on the Gilbert damping in comparison to the STT switching[24-26]. As the Gilbert damping is usually much smaller than the unity, this insensitivity to the damping makes the write current of SOT switching high[24-26]. The orthogonal configuration also demands a symmetry-breaking field to achieve the deterministic switching because the SOT tends to align the magnetization in the $y$ direction, not in the $z$ direction.

Recent studies found that it is possible to rotate the spin polarization from the $y$ direction by introducing an additional FM: FM1 (free layer)/NM/FM2 trilayers. The anomalous Hall effect of FM2[27,28] generates a spin current polarized in $\hat{\mathbf{m}}_2$, i.e., magnetization direction of FM2. The interface-generated spin currents at the NM/FM2 interface[23,29-31] generates a spin current with a spin polarization in $(\hat{\mathbf{m}}_2 \times \hat{\mathbf{y}})$ through the spin-orbit precession process. Therefore, a spin current created in the trilayers can have an additional spin-polarization component in the $z$ direction. This additional spin-$z$ spin current in the CIP geometry naturally allows field-free deterministic switching of perpendicular magnetization as recently demonstrated in an experiment[23]. It is expected that the write current would decrease due to the additional spin-$z$ spin current[27], but the exact expression of switching current in the presence of additional spin-$z$ spin current has not been investigated yet.

In this work, we theoretically and numerically investigate the switching current required for perpendicular magnetization switching induced by a spin current with an arbitrary spin polarization direction. Our main purpose is to provide the analytic expression of the switching current, which can be used as a design rule for SOT-MRAMs based on the aforementioned ferromagnetic trilayers. As the spin polarization direction is different depending on the mechanism, we do not focus on a specific mechanism but investigate the effect of arbitrary spin polarization directions.

**Analytical analysis**

Magnetization dynamics driven by a spin current with an arbitrary spin polarization direction is described by the Landau-Lifshitz-Gilbert equation including the both damping-like torque (DLT) and field-like torque (FLT) as,

$$\frac{d\hat{\mathbf{m}}}{dt} = -\gamma \hat{\mathbf{m}} \times \mathbf{H}_{\text{eff}} + \alpha \hat{\mathbf{m}} \times \frac{d\hat{\mathbf{m}}}{dt} + \gamma c_{j,D} \hat{\mathbf{m}} \times (\hat{\mathbf{m}} \times \hat{\boldsymbol{\sigma}}) + \gamma c_{j,F} \hat{\mathbf{m}} \times \hat{\boldsymbol{\sigma}}, \qquad (1)$$

where $\hat{\mathbf{m}}$ is the unit vector along the magnetization of FM1, $\hat{\boldsymbol{\sigma}}$ is the unit vector along the spin polarization, $\gamma$ is the gyromagnetic ratio, $\mathbf{H}_{\text{eff}}$ is the effective uniaxial anisotropy field $H_{K,eff} = 2K_{eff}/M_S$ in the $z$ direction, $\alpha$ is the damping constant, $c_{j,D(F)} = (\hbar \theta_{D(F)} J / 2 e M_s t_z)$ is the magnitude of DLT(FLT), $\theta_{D(F)}$ is the effective DLT(FLT) efficiency, $J$ is the charge current density flowing in the plane (along the $x$ axis), $e$ is the electron charge, $M_S$ is the saturation magnetization, and $t_z$ is the thickness of FM1. We assume that $\hat{\sigma} = (0, \cos\eta, \sin\eta)$ is a spin polarization direction, because the system is cylindrical symmetry in the *x-y* plane, and $\eta$ represents the spin-polarization angle. We express the magnetization vector as $\hat{\mathbf{m}} = (\cos\phi\sin\theta, \sin\phi\sin\theta, \cos\theta)$, where $\theta$ ($0 \leq \theta \leq \pi$) is the polar angle and $\phi$ ($0 \leq \phi < 2\pi$) is the azimuthal angle. In order to derive analytic expressions of the switching current, we ignore FLT because it induces magnetization precession, which complicates magnetization dynamics[26,32]. We note that the effect of FLT on the switching current is insignificant when $\hat{\sigma}$ has a sizable z component, which will be verified numerically below.

For a charge current density smaller than a switching current density, Eq. (1) has a static solution with the equilibrium tilting angles $\theta_{eq}$ and $\phi_{eq}$ satisfying:

$$c_{j,D} \cos\eta \cos\phi_{eq} - H_{K,eff} \cos\theta_{eq} \sin\theta_{eq} = 0, \qquad (2)$$

$$\sin\eta \sin\theta_{eq} - \cos\eta \cos\theta_{eq} \sin\phi_{eq} = 0. \qquad (3)$$

Depending on $\eta$, switching conditions can be classified into two cases. The first case is the instability condition, corresponding to no solutions of $\theta_{eq}$ and $\phi_{eq}$ satisfying Eqs. (2) and (3). By combining Eqs. (2) and (3), we obtain the switching current density $J_{sw,1}$ and tilting angles $(\theta_{sw,1}, \phi_{sw,1})$ for the instability condition as

$$J_{sw,1} = \frac{H_{K,eff} M_s e t_z}{\hbar \theta_D \cos\eta} \frac{\sin 2\theta_{sw,1}}{\sqrt{1-(\tan\eta \tan\theta_{sw,1})^2}} = \frac{2 K_{eff} e t_z}{\hbar \theta_D \cos\eta} \frac{\sin 2\theta_{sw,1}}{\sqrt{1-(\tan\eta \tan\theta_{sw,1})^2}}, \qquad (4)$$

$$\theta_{sw,1} = \cos^{-1}\left(\frac{\sqrt{4-3\cos^2\eta+\cos\eta\sqrt{9\cos^2\eta-8}}}{2}\right), \quad \phi_{sw,1} = \sin^{-1}(\tan\eta \tan\theta_{sw,1}). \qquad (5)$$

For $\eta = 0$ (thus, $\hat{\sigma} = \hat{y}$), Eqs. (4) and (5) are simplified as

$$J_{sw,1} = \frac{M_s e t_z}{\hbar \theta_D} H_{K,eff} = \frac{2K_{eff} e t_z}{\hbar \theta_D}, \quad \theta_{sw,1} = \pi/4, \quad \phi_{sw,1} = 0, \tag{6}$$

which is consistent (except for the in-plane external field) with our previous result[24].

The second case is the anti-damping condition. In this case, the switching occurs when the DLT overcomes the intrinsic damping torque. Because the SOT directly competes with the damping torque, the magnetization switching occurs through many precessions as for the conventional STT switching. As a result, the switching current can be obtained for the condition that the precession angle becomes larger with time evolution. After rotating the coordinate system to the magnetization tilted by SOT, we use the spin-wave ansatz[33] of $\hat{\mathbf{m}} = (m_{x'} e^{i\omega t}, m_{y'} e^{i\omega t}, 1)$, where $(|m_{x'}|^2, |m_{y'}|^2) \ll 1$ (here prime means the rotated coordinate), and obtain an equation satisfying the condition that intrinsic damping and SOT are cancelled out (equivalently, the imaginary part of spin-wave dispersion vanishes), given as

$$\alpha H_{K,eff}(1 + 3\cos 2\theta) = 4c_{j,D}(\cos\theta \sin\eta + \cos\eta \sin\theta \sin\phi). \tag{7}$$

For the second case, one can obtain the expressions for the switching current density $J_{sw,2}$ and tilting angles $(\theta_{sw,2}, \phi_{sw,2})$ by combining Eqs. (2), (3), and (7). However, the expressions are too long to be presented in the paper. Instead, we show simplified analytic expressions with the assumption of $\phi_{sw,2} \approx 0$, which is reasonable for most ranges of $\eta$ as shown below. The simplified expressions are:

$$J_{sw,2} \approx \frac{2AB H_{K,eff} M_s e t_z \sec\eta \tan\eta}{9\alpha \hbar \theta_D} = \frac{4AB K_{eff} e t_z \sec\eta \tan\eta}{9\alpha \hbar \theta_D}, \tag{8}$$

$$\theta_{sw,2} \approx \tan^{-1}\left(\frac{A \tan\eta}{B\alpha}\right), \quad \phi_{sw,2} \approx \tan\theta_{sw,2} \tan\eta, \tag{9}$$

where $A = -1 + \sqrt{1 + 6\alpha^2 \cot^2\eta}$, $B = \sqrt{3 + \frac{12}{1+\sqrt{1+6\alpha^2 \cot^2\eta}}}$. When $\eta = \pi/2$, $J_{sw,2}$ is

obtained by taking the limit of $\eta \to \pi/2$, given as

$$J_{sw,2} \approx \alpha \frac{2e}{\hbar} \frac{M_s t_z H_{K,eff}}{\theta_D} = \alpha \frac{4e}{\hbar} \frac{t_z K_{eff}}{\theta_D}, \quad (10)$$

which is consistent with the switching current density for STT switching.

**Numerical Results**

In order to check the validity and applicability of the above analytic expressions, we perform macrospin simulation by numerically solving Eq. (1). We use following modeling parameters: area of free layer = 900 nm$^2$, ferromagnet thickness $t_z = 1$ nm, gyromagnetic ratio $\gamma = 1.76 \times 10^7$ Oe$^{-1}$s$^{-1}$, effective perpendicular anisotropy constant $K_{eff} = 2 \times 10^6$ erg/cm$^3$, saturation magnetization $M_s = 1000$ emu/cm$^3$, Gilbert damping $\alpha = 0.005$, effective DLT efficiency $\theta_D = 0.3$, effective FLT efficiency $\theta_F = 0$, external magnetic field $H_x = 300$ Oe only for $\eta = 0$ ($\hat{\sigma} = \hat{y}$), current pulse-width $\tau = 200$ ns, and rise/fall time = 0.2 ns.

Figure 1 shows the switching current density ($J_{sw}$) and tilting angles ($\theta_{sw}, \phi_{sw}$) as a function of $\eta$ and time evolution of $\hat{m}$. Numerical results (symbols) are in agreement with the analytic solutions (lines). As shown in Fig. 1(a) and its inset, numerically obtained $J_{sw}$ is inconsistent with Eq. (5) (i.e., the instability condition) but is consistent with Eq. (8) (i.e., the anti-damping condition) in wide $\eta$ ranges except for small $\eta$. The good agreement between numerically obtained $J_{sw}$ and Eq. (8) justifies the assumption of $\phi_{sw,2} \approx 0$ in wide $\eta$ ranges, which is also seen in Fig. 1(b). Fig. 1(c)-(e) show time evolution of $\hat{m}$ for different $\eta$. When $\eta = 0.002$ [Fig. 1(c)], time evolution of $\hat{m}$ is similar with conventional SOT switching except that the deterministic switching is achieved without an external field. When $\eta = 0.05$ [Fig. 1(d)] and $\eta = 0.2$ [Fig. 1(e)], $\hat{\mathbf{m}}$ first rotates around the tilted axis, which is similar to the conventional STT switching. After the precession angle reaches a specific value, $\hat{\mathbf{m}}$ stays in a direction tilted from $-z$ direction while the current is applied [Fig. 1(d) and inset of Fig. 1(e)]. The amount of tilting from $-z$ direction depends on $\eta$ and applied current. For all $\eta$ ranges except for $\eta = \pi/2$, $\hat{\mathbf{m}}$ is aligned with $-z$ direction only after the current is turned off.

The results shown in Fig. 1 indicate that the switching condition changes from the instability condition to the anti-damping condition as $\eta$ (equivalently, the $z$ component of spin polarization) increases. This $\eta$ dependence of $J_{sw}$ can be understood as follows. $J_{sw}$ is determined by $\min[J_{sw,1}, J_{sw,2}]$. In the small $\alpha$ limit, $J_{sw,2}$ is approximated as

$$J_{sw,2} \approx \frac{\alpha}{\sin\eta} \frac{2e}{\hbar} \frac{M_s t_z}{\theta_D} H_{K,eff}, \tag{10}$$

leading to $J_{sw,2}/J_{sw,1} \approx 2\alpha/\sin\eta$. Therefore, for $2\alpha/\sin\eta < 1$, the switching is governed by the anti-damping condition, whereas, for $2\alpha/\sin\eta > 1$, the switching is governed by the instability condition. This analysis also sets an approximated critical spin-polarization angle $\eta_c = \sin^{-1} 2\alpha$, above (below) which the switching is governed by the anti-damping (instability) condition.

From above results, one finds that $J_{sw}$ becomes small as $\eta$ increases (i.e., spin-$z$ component increases)[27], confirming a possibility to resolve the second issue, i.e., high write current for conventional SOT switching. To address this possibility in more detail, we show material parameter and current pulse-width ($\tau$) dependences of $J_{sw}$. Figure 2 shows dependences of $J_{sw}$ on (a) damping constant, (b) effective anisotropy constant, (c) saturation magnetization, and (d) current pulse-width. Increased damping [Fig. 2(a)] or increased anisotropy [Fig. 2(b)] increases $J_{sw}$, as expected from Eq. (8). In contrast, the saturation magnetization does not affect $J_{sw}$ [Fig. 2(c)], which is also consistent with Eq. (8). A result that is not captured by Eq. (8) is the pulse-width dependence $J_{sw}$ [Fig. 2(d)] : $J_{sw}$ increases with decreasing the pulse-width. This increased $J_{sw}$ at a short current pulse is understood by the fact that the switching occurs through many precessions, which increase the time duration to escape the energy minimum. The results shown in Fig. 2 suggest that $\eta$ close to $\pi/2$ (or, equivalently, a large $z$ component of spin polarization) and a small damping are two preconditions to reduce $J_{sw}$. Even though $J_{sw}$ also reduces with decreasing the anisotropy, it is not a free parameter to maintain a long retention time for non-volatile applications.

We also numerically study how the FLT and thermal fluctuation affect the switching current. We perform macrospin simulation including Gaussian-distributed random thermal fluctuation fields (mean = 0, standard deviation = $\sqrt{2\alpha k_B T/(\gamma M_s V \delta t)}$, where $\delta t$ is the integration time step[34]). We assume that the temperature is 300 K, corresponding to the energy barrier $\Delta \approx 43.5$ for our parameter set. We repeat simulations 1000 times for each current density to consider the randomness of thermal fluctuation.

Figure 3(a) shows $J_{sw}$ as a function of $\eta$ for different FLT/DLT ($\theta_F/\theta_D$) ratios. We find that $J_{sw}$ exhibit clearly different dependences on $\eta$ between small $\eta$ ($\eta < 0.2$) and large $\eta$ ($0.2 < \eta$) ranges. In small $\eta$ ranges, non-deterministic switching occurs when the sign of FLT is opposite to that of DLT [in our sign conventions; see Eq. (1)][26,32]. For the case where the sign of FLT is same with that of DLT, $J_{sw}$ is high in comparison with that with $\theta_F = 0$. In large $\eta$ ranges, FLT does not significantly affect $J_{sw}$. This result means that a large $\eta$ (equivalently, large spin-$z$ component of spin current) allows for low $J_{sw}$ and deterministic switching simultaneously regardless of the FLT. Figure 3(b) shows switching probability ($P_{sw}$) for $\tau = 5$ ns and different spin polarization angles ($\eta$) as a function of the current density. One finds that $J_{sw}$ decreases with increasing $\eta$, consistent with the above results. For small $\eta$ ($\eta = 0.002$ and $0.02$), deterministic switching does not occur due to thermal fluctuation. We also find that, for the parameter set we used, $\eta$ larger than 0.1 is required for deterministic switching. Figure 3(c) shows switching current ($I_{sw}$) as a function of $\tau$ at various $\eta$. Here, we compare $I_{sw}$ for the case where $\eta \geq 0.2$. $I_{sw}$ is obtained from $J_{sw}$ at $P_{sw} = 1/2$, multiplied by a cross section area, normal to the current-flow direction. For CIP case, we assume that the cross section area $A_{CIP}$ is 150 nm² (= 30 nm × 5 nm). For a comparison, we also plot $I_{sw}$ of the conventional spin-transfer torque (STT) switching for the spin polarization $P$ of 0.3. For STT switching, we use the cross section area $A_{CPP}$ of 900 nm² because it is the CPP geometry and thus must be the same as that of free layer. Here we compare $I_{sw}$, instead of $J_{sw}$. The reason is that $I_{sw}$ is more relevant to device applications than $J_{sw}$, because $I_{sw}$, not $J_{sw}$, determines the transistor size and thus the device scalability.

The most important observation from Fig. 3(c) is that $I_{sw}$ for SOT with a tilted spin polarization is smaller than that for STT even at $\tau$ of 1 ns. Using the approximate solution [Eq. (10)] for SOT switching and $I_{sw}$ for the conventional CPP STT switching[24], the ratio $I_{sw}(\mathrm{STT})/I_{sw}(\mathrm{SOT})$ is given by

$$\frac{I_{sw}(STT)}{I_{sw}(SOT)} = \frac{A_{CPP}}{A_{CIP}} \frac{\theta_D \sin\eta}{P}. \tag{11}$$

As $A_{CPP}/A_{CIP}$ is about a factor of 5 for 30 nm MRAM cell, $I_{sw}(\mathrm{SOT})$ is smaller than $I_{sw}(\mathrm{STT})$ when $\theta_{SH} \sin\eta > 0.2P$. This result shows that the SOT with a tilted spin polarization is able to reduce the switching current below those of not only conventional SOT switching but also conventional STT switching.

**Discussion**

In conclusion, we theoretically and numerically investigate the switching current for SOT switching of perpendicular magnetization in ferromagnetic trilayers. We confirm that the spin-$z$ component of spin polarization, originating from either the anomalous Hall effect[27,28] or the interfacial spin-orbit coupling effect[23,29-31], enables the deterministic switching at a low current. This practically attractive consequences from the tilted spin polarization will be beneficial for SOT memory and logic devices operated at low power. Moreover, analytical expressions of the switching current for an arbitrary spin polarization can be used as a guideline to design SOT devices and also to interpret experimental switching data obtained for unconventional spin currents of which spin polarization is deviated from the $y$ direction by known and yet-unknown mechanisms.

**Acknowledgement**

This work was supported by the National Research Foundation of Korea (Grant No. NRF-2015M3D1A1070465) and Samsung Electronics.

**Author contributions**

K.-J. Lee conceived and supervised the study. D.-K. Lee developed an analytical model and performed macrospin simulations. All the authors wrote the manuscript.

**Additional information**

Competing interests: The authors declare no competing interests.

**Figure Captions**

**Figure 1. Switching properties induced by an arbitrary spin polarization direction ($\vec{\sigma} = cos\eta \hat{y} + sin\eta \hat{z}$).** (a) Switching current density ($J_{sw}$) and (b) switching tilting angles ($\theta_{sw}, \phi_{sw}$) as a function of $\eta$. Inset of (a) is $J_{sw}$ at the $\eta$ range from 0 to 0.01 radian. Symbols are macrospin simulation results and lines are analytic expressions. Time evolution of $\hat{\mathbf{m}}$ when (c) $\eta = 0.002$, (d) $\eta = 0.05$, and (e) $\eta = 0.2$. Inset of (e) is trajectory of $m_z$ when the current is turned off.

**Figure 2. Material parameter and current pulse-width dependences of $J_{sw}$.** Switching current density ($J_{sw}$) depending on (a) damping constant, (b) effective anisotropy constant, (c) saturation magnetization, and (d) current pulse-width ($\tau$) as a function of $\eta$. Symbols are macrospin simulation

results and lines are analytic solutions.

**Figure 3. Switching properties obtained from the macrospin simulation including FLT and thermal fluctuation fields.** (a) Switching current density $J_{sw}$ for different FLT/DLT ratios as a function of $\eta$. (b) Switching probability curves for current pulse-width $\tau = 5$ ns and different spin polarization angles as a function of the current density. (c) Switching current $I_{sw}$ for different spin polarization angles as a function of current pulse-width.

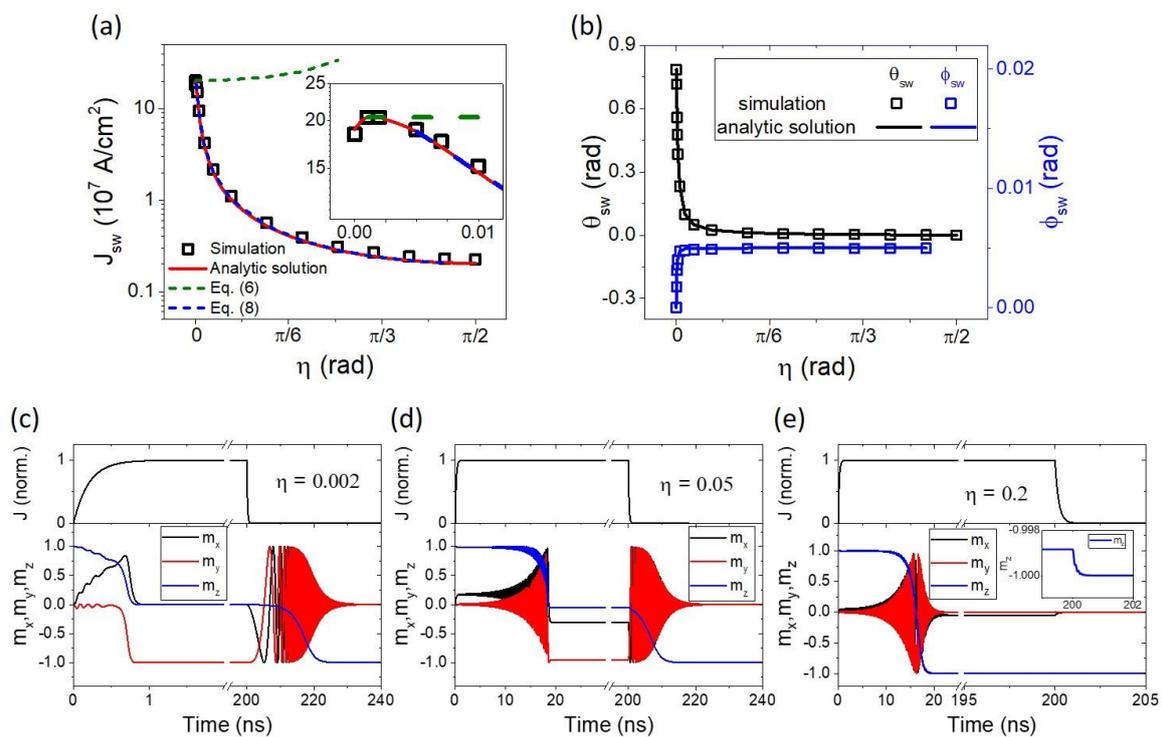

Fig. 1.

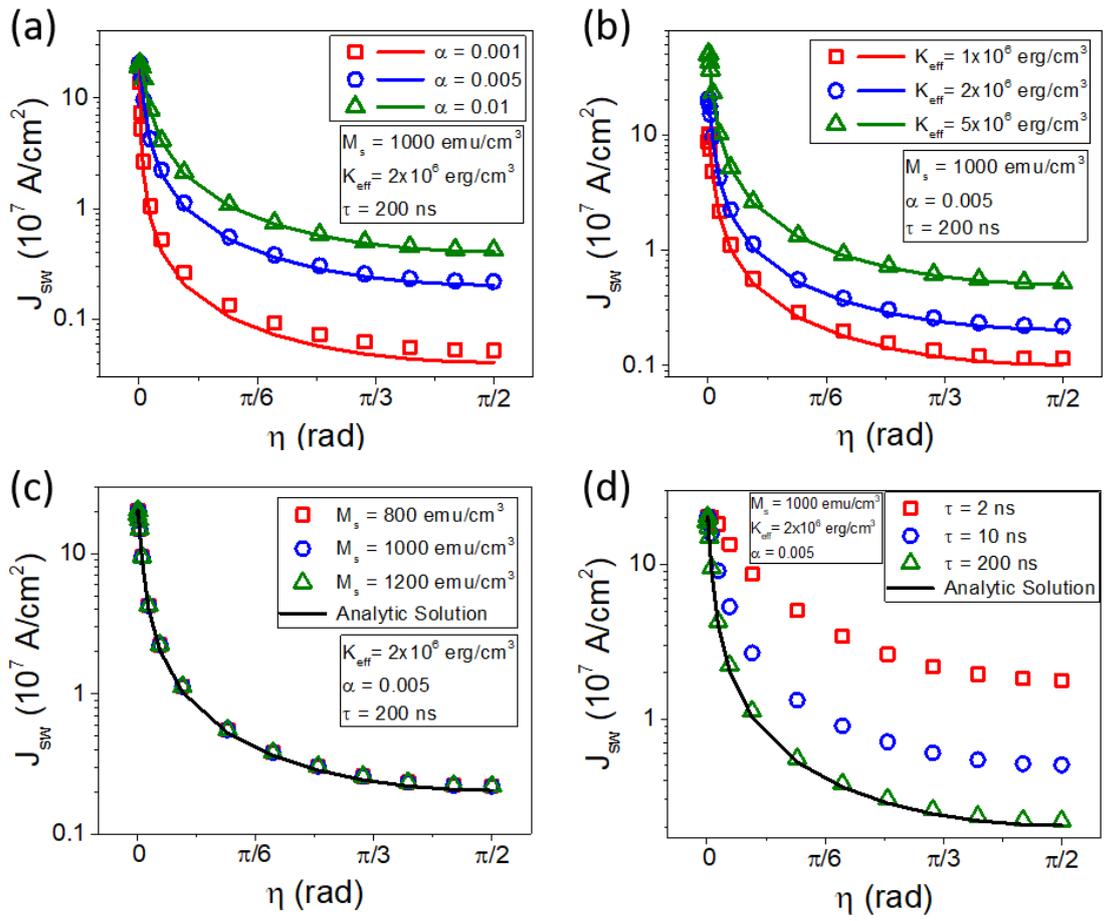

Fig. 2.

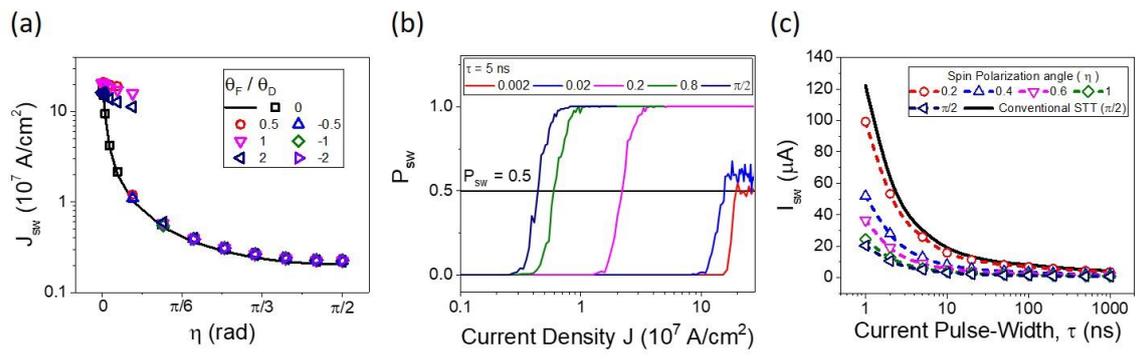

Fig. 3.